\documentstyle[prb,aps,multicol,epsfig]{revtex}
\begin{document}
\newcommand{\mex}[1]{$ \langle #1 \rangle $ }
\twocolumn[\hsize\textwidth\columnwidth\hsize\csname
@twocolumnfalse\endcsname
\draft

\title{
Generalized stacking fault energetics and dislocation properties:
compact vs. spread  unit dislocation structures in  TiAl and CuAu. }

\author{
Oleg N. Mryasov$^{1}$, Yu. N. Gornostyrev$^{1,2}$ and A.J. Freeman$^{1}$
}
\address{
\(^1\)  Department
of Physics and Astronomy, Northwestern University Evanston, IL
60208-3112. \\
\(^2\) Institute of Metal Physics, Ekaterinburg, Russia.
}
\maketitle

\begin{abstract}
 We present  a
 general scheme for analyzing the  structure and mobility of   dislocations
based on solutions of the  Peierls-Nabarro
model with  a two component displacement field   and
restoring forces determined
from the ab-initio generalized stacking fault  energetics
(ie., the so-called $\gamma$-surface).
The approach is  used to investigate
dislocations in L1$_{0}$ TiAl and CuAu;
predicted differences in the unit  dislocation properties
are explicitly related with  features
of the $\gamma$-surface geometry.
A unified description of  compact, spread and split dislocation cores is
provided
with an important characteristic
"dissociation path" revealed by this  
 highly tractable scheme.
\end{abstract}

\vskip2pc]
\section{Introduction}

   In the development of modern materials
with advanced high-temperature structural
properties \cite{pope_interm,tial_rev},
there is a growing awareness of
the  importance of a fundamental
understanding of  the role of specific lattice defects.
One such defect,  dislocations,
is  always present in most  real materials and influences  a
variety of properties, including
mechanical, electric, optical and magnetic.
High-temperature intermetallics is one of the most vivid examples where
particular dislocations  control deformation, fracture and other properties that are 
most important in terms of  their technological applications.

Dislocations are a unique example of the extended defect for which  a single
length scale description (atomistic or continuum)
 is not quite appropriate - a feature that is one of the main sources
of difficulties in their  theoretical description. 
There are hundreds
of atoms in a dislocation
core (requiring an atomistic treatment) and thousands of atoms are  involved in
the formation of the elastic dislacements fields (where a continuum description is more
appropriate) so that  neither entirely atomistic nor continuum
descriptions are suitable to the problem but rather need to be combined.

Classical  atomistic
simulations, however, proved to be very useful and provided important insights
into  their fundamental properties  in metals \cite{vitek_md}.
To make use of atomistic simulations for modeling dislocations in particular
material reliable model of interatomic interaction has to be developed.
This, however, poses major principal problem for the use of classical atomistic
simulations.
Unfortunately, in  most of interesting cases
(including intermetallics like  L1$_{0}$ TiAl)
 interatomic interactions are  very complex and 
so their satisfactory description can not be guaranteed withing 
existing central force models.
This  significantly limits  abilities of classical atomistic simulations
for predictive analysis 
dislocations properties particularly in intermetallic alloys
with their complex character of interatomic interactions
\cite{vitek_direct_bond}.
%
%
To date, the few quantum molecular dynamic simulations  of dislocations
have been restricted to s,p systems (Si, SiC), employing periodic boundary 
conditions with a few hundred atoms in the supercell \cite{abindisl}.
This approach remains  unexceptably  expensive and  
has principal difficulties to be applied  for  intermetallics containing
3d-elements.

The complexity of the problem  has
stimulated  renewed  interest \cite{duePRB,ourPN1,duePRL} in the simple
but tractable
Peierls-Nabarro (PN) dislocation model,
 which incorporates atomistic effects in an approximate  way  into a continuum
framework \cite{PN}.
In this model,  the effects of  
lattice 
discreteness are confined to
a slip plane  with  a local relation
for non-linear atomistic restoring
forces which can be determined from  an effective rigid shift interplanar
potential known
as a generalized stacking fault (GSF) energy 
\cite{vitek_md}.
The PN model proved to be very useful and
has strongly   influenced  conceptually the theory of dislocations, fracture
and related mechanical properties \cite{vitek_md}.
Naturaly appeared  in PN model,
the energy of the relative displacements of two halves of the crystal
(ie., the so-called generalized stacking fault (GSF) energy,  or
$\gamma$-surface)
was introduced  and studied as an important general
characteristic of the mechanical
response in solids \cite{vitek_md,JRRice:92}.
%
 
 Despite its  well-known  approximations
\cite{HirthLote,bulatov,Belts,Nabarro97}, the PN model  is  very
 attractive since
it offers a tractable description with parameters ($\gamma$-surface)
accessible by highly accurate quantum mechanical calculations
\cite{ourPN1,duePRL}. As was recently demonstrated, the  applicability of  
the Peierls model framework
can be extended in a semi-discrete version even for  such  an extreme case as  narrow
dislocation cores in Si \cite{bulatov}.
In this work, the classic Peierls model  dislocation energy 
functional 
 is minimized  using a discrete scheme
which   allows    avoiding  assumptions  
of the dislocation core planarity and  the Nabarro approximation for calculating Peierls stresses.  
%
Unfortunately, the proposed semi-discrete scheme   is purely numerical;
however,  in the theory of such complex lattice defects as dislocations  it is
highly desirable to keep the  high tractability of the theoretical approach as much as possible.
 
To date, the PN model  analysis  with  first-principles  parametrization 
for  GSF energetics was restricted, however, to considering dislocations with
a single  component displacement field (one dimensional (1D) PN model )  in Si
\cite{duePRB}
and in B2  intermetallic compounds \cite{ourPN1}.
The abilities of the PN model  to deal with  dislocation cores in intermetallics
has not yet been studied  systematically, except for our earlier overall
encouraging experience with the 1D-PN model for  dislocations in B2 NiAl and FeAl \cite{ourPN1}.
However, the  two component displacement field (two-dimensional (2D) PN model)
with the possibility of  dislocation  dissociation
is  quite typical situation in general and,  in particular, is necessary to
be  considered in the  case of  unit  dislocations for
L1$_{0}$  TiAl and CuAu, materials with similar  fcc-like structure but very
different mechanical behaviour.

In contrast with the 1D PN model, exact solutions of the 2D-PN model are unknown 
even  for  the  simplest sinusoidal restoring force law.
Recently, Schoeck presented a direct variational method 
for analyzing, on the energy functional level, the 2D-PN model using ansatz solutions
with geometric parameters \cite{schoeck_94}. This method
 has been used   to  calculate dislocation core structures
 corresponding to  model  GSF surfaces \cite{schoeck_98}.
Here we present an   alternative  scheme for solving 2D-PN model equations 
within  a wide class of analytic functions  which makes 
this scheme unified (i.e., free of parameters and assumptions on dislocation 
structure), suitable for analyzing realistic/complex  $\gamma$-surfaces
and physically transparent.   
We apply this scheme   to  analyze unit dislocation structure and mobility
in L1$_{0}$  TiAl and CuAu alloys with GSF energetics determined
from ab-initio calculations.

 In this work, we demonstrate how  advantages of the PN model's high
tractability can be fully revealed due to the use of a general and physically transparent
scheme for analyzing  solutions in  the generalized case of a two component 
displacement field (2D)  and with a first-principles determination of
the $\gamma$-surface energetics. 
We focus on an analysis of the unit dislocation structure in  
L1$_{0}$ TiAl and CuAu  using the proposed scheme since
this type of dislocation is  important for
understanding  these materials with very different mechanical behaviour\cite{tial_rev}. 
We demonstrate that this proposed  approach  explains
qualitative differences in dislocation structure and mobility in a very natural way and 
establishes an  explicit  relation between $\gamma$-surface geometry and   
dislocation properties  within the 2D PN model. This allows one to relate  
GSF energetics  determined from ab-initio electronic structure 
calculations with  dislocation structure using a unified and highly transparent  procedure. 

\section{Approach and Method}

The structure of dislocations within the  2D PN model is described
from a balance between elastic 
 and atomistic restoring forces
which is expressed in  a system of integro-differential equations
\cite{seeger_53,schoeck_94}

\begin{eqnarray}
\frac{K_{\alpha \beta}}{2\pi} \int_{-\infty }^{\infty} \frac{d\xi }{\xi -x}
\frac{%
d u_{\beta}(\xi )}{d \xi }=F_{\alpha}(\vec{u}) %
\label{eqn2dpn}
\end{eqnarray}
Here $\beta$ is  the index for components of the  dislacement $\vec{u}$, with 
$\beta$=1  for screw  and $\beta$=2 for  edge,   and
 $K_{\alpha \beta}=\mu D_{\beta}\delta_{\alpha \beta}$ \cite{note1}, 
$D_{\beta}$ is a parameter equal to 1 for  $\beta=1 $
and 1/(1-$\nu $) for $\beta$=2. There are at least two
difficulties to applying  the PN model for realistic situations: (i)
the restoring force law is unknown in general  and (ii)
an analysis of Eq. (\ref{eqn2dpn}) seems to  be difficult for an arbitrary
restoring force law. We follow the   GSF concept \cite{vitek_md}  and define
the restoring force in the local approximation \cite{Belts}
as  $ F_{\beta}(\vec{u})=-d E(\vec{u}) / d u_{\beta} $,  where
GSF energies $E(\vec{u})$ are calculated 
for experimental lattice structure parameters 
employing  the local  density full-potential linear muffin-tin orbital (FLMTO) band
structure technique \cite{Meth-prb-88}  
with the  Ceperly-Alder form for the exchange-correlation potential.

In fact, special points on the $\gamma$-surface known as planar stacking
fault energies (APB, ISF, CISF etc.)  have been calculated  by various  ab-initio methods
for  a number  of  metals, alloys and compounds (see references in \cite{PaxtonBook}). 
Thus, calculations of the entire $\gamma$-surface do not pose principal difficulties
but may be rather challenging from a numerical stand point
if one would like to achieve, as much as possible, equally high  accuracy  for all $\gamma$-surface sampling
 points.
Most of the  numerical aspects one may  face in  ab-initio calculations of the GSF energetics
have been    summarized by   Paxton \cite{PaxtonBook}.
Probably, the first example of  the entire $\gamma$-surface determined using ab-initio techniques is
a pseudopotential plane wave (PPW)  calculation for Si  by Kaxiras and Duesbery \cite{kax-93}.
The PPW approach, however,  is somewhat less natural to use 
if rather localized  d and f electons are present in a system. Thus, in this work
we make use of the  all-electron FLMTO technique   which allows one   to treat equally well 
both localized and delocalized electronic states with reasonable accuracy  and  high numerical 
efficiency\cite{Meth-prb-88}. 

One should emphasize  here  that only  in the local approximation
is so simple a relation  provided  with GSF energies and hence restoring forces 
are so easily accessible  by ab-initio  techniques.
As demonstrated by  Miller and Phillips \cite{Belts} using 
MD simulations, the local
approximation merely affects  dislocation structure but may  
be important for calculating  Peierls stresses.
Any corrections beyond the local approximation  involve a much more
complex procedure \cite{Belts} and to the best of our knowledge have not  yet been  realized
using ab-initio calculations.

An analysis of the solutions of Eq.(\ref{eqn2dpn})  is  rather complex
from a mathematical point of view.
An elegant idea proposed by 
Lejeck  for    finding solutions of the   1D PN model equations 
in a wide class of analytic functions \cite{Lejcek} seems to be general and   suitable
for cases of complex/realistic  restoring force laws. 
Details of such an analysis for  the 1D PN model with generalized 
restoring forces  determined for NiAl  from ab-initio calculations   are discussed
in our earlier work \cite{ourPN1}.  
Our results  for this realistic example demonstrate that the scheme for solving the 
PN model type equations
based  on Lejeck's idea can be 
very convenient and  provides the  accuracy and numerical stability desired.
In the 2D PN model case, however, additional complications related with the   
coupling of equations (cf. Eq.(\ref{eqn2dpn})) have to be overcome.
In this work,  we  present a scheme for solving the 2D PN model  
 equations with an arbitrary restoring force law 
   which is  based on  Lejcek's  idea \cite{Lejcek} as generalized  for  the 2D case.

Solutions of
Eq. (\ref{eqn2dpn})  are found  by constructing an analytic
complex function G$_{\beta}(x)=\rho_{\beta}(x)+i \cdot \frac{2}{\mu D_{\beta}}
F_{\beta}(x) $
where $\rho_{\beta}(x) = du_{\beta}(x)/dx $  is the  dislocation density,
and $ F_{\beta}(\vec{u}(x)) $ the  restoring force
and  are given as expansions
\begin{eqnarray}
\rho _{\beta}(x) = Re \sum_{k=1}^N \sum_{n=1}^{p_k} \frac{A_{nk}^{\beta}}
{(x-z_k^{\beta})^n},
\label{RhoEqn}
\end{eqnarray}
\begin{eqnarray}
\frac 2 {\mu D_{\beta}} F_{\beta}(x)=
-Im \sum_{k=1}^N \sum_{n=1}^{p_k} \frac{A_{nk}^{\beta}}{(x-z_k^{\beta})^n}
\label{ForEqn}
\end{eqnarray}
Here $N$ is the number of  poles with maximal  order $p_k$ at  points
in the complex plane, $z_k^{\beta}=x_k^{\beta}+i\zeta _k^{\beta}$
($k=1,...,N$).
This representation is not only mathematically justified
(the  structure of  the PN model equations is analogous to the  Hilbert
transform)
but is also physically transparent. Indeed,  the number of poles in the
expansion  for
 $\rho _{\beta}(x)$ determines the  number of   
 cores 
 (partial or fractional dislocations)
 with a width $\zeta _k^{\beta}$ and positions of the centers, $x_k^{\beta}$.
The solutions,  $ u_{\beta} (x) $, can be found by integrating $\rho
_{\beta}(x)$ with
appropriate boundary conditions. Then, for an fcc lattice and a screw
orientation
of the unit  dislocation in the \{111\} plane, the functions $ u_{\beta}(x) $
and $ F_{\beta}(\vec{u}(x)) $ for p$_{k}$=2 and N=2 can be written in the  convenient
parametric form

\begin{eqnarray}
\frac {2 \pi u_1}{b} = \frac 1 2
\left[\theta _1^1 + \theta _2^1 -\frac {\alpha _1 -1}{\alpha _1}
(sin\theta _1^1 + sin\theta _2^1)\right]
\label{disSEqn}
\end{eqnarray}

\begin{eqnarray}
\frac {\pi u_2}{b_e} = \frac 1 2
\left[\theta _1^2 - \theta _2^2 -\frac {\alpha _2 -1}{\alpha _2}
(sin\theta _1^2 - sin\theta _2^2)\right]
\label{disEEqn}
\end{eqnarray}

\begin{eqnarray}
\frac {4 \pi \omega _0^{1}}{\mu D_1 b}F_1 & = & - \frac 1 {2\alpha _1}
\left[sin\theta _1^1 + sin\theta _2^1  + \frac {2(\alpha _1 -1)}{\alpha _1}
\times \right.
\nonumber \\
&  \times &
\left.
(sin^2(\theta _1^1/2)sin \theta _1^1 +
 sin^2(\theta _2^1/2)sin\theta _2^1) \right]
\label{forsSEqn}
\end{eqnarray}

\begin{eqnarray}
\frac {2 \pi \omega _0^{2}}{\mu D_2 b_e}F_2 &  = & - \frac 1 {2\alpha _2}
\left[sin\theta _1^2 - sin\theta _2^2 + \frac {2(\alpha _2 -1)}{\alpha _2}
\times \right.
\nonumber \\
&  \times &
\left.
(sin^2(\theta _1^2/2)sin \theta _1^2 - sin^2(\theta _2^2/2)sin\theta _2^2)
\right]
\label{forsEEqn}
\end{eqnarray}
Here $\theta_k^{\beta}=2arc cot((x - x_{k}^{\beta })/\zeta^{\beta}$),
$\zeta^{\beta}=\alpha_{\beta }\omega_0^{\beta}$, where
$\omega _0^{\beta }$ is the width of the  dislocation core which appears
in  the solution of the original PN model for a sinusoidal restoring force law,
$b$ is the  total  Burgers vector and
$b_e$ is the  edge component of the partial dislocation Burgers vector.
Thus, the complete set of dislocation  structure parameters,
$\alpha_{\beta}$ and  $ x_{k}^{\beta } $ 
(including possible splitting) 
can be determined from the general restoring force law
\cite{note3}, $F_{\beta} (\vec{u})$.

However, there is a complication related with
an essential feature of the 2D PN model.
Now,  the  2D PN model  equations (Eq.(\ref{eqn2dpn}))
are  coupled only \cite{note1}
through the restoring force which is a function of the unknown a priori 
   "dissociation path"  determined
   in the two component displacement space 
and given
by the dependence 
$u_{2}=f(u_{1})$ of the edge ($u_{2}$) on the screw ($u_{1}$)
component.
 Thus,  function $f(u_{1})$ has to be found along with
solutions of Eq.(\ref{eqn2dpn}).
Hence, we propose an iterative procedure which starts with calculations of the
restoring forces,
$F_{\beta}(u_1, u_2)$,  and the corresponding dislocation structure
parameters for a linear "dissociation path" \cite{note4}, $u_{2}=K \cdot
u_{1}$. Then,
as  convergence is achieved,  both the actual "dissociation path" and the
solutions $\vec{u}(x)$  can
be determined. 

\section{Results and Discussion}
\label{sec:results}

\begin{figure}[!htb]
\vskip  0cm
\centerline{
\psfig{figure=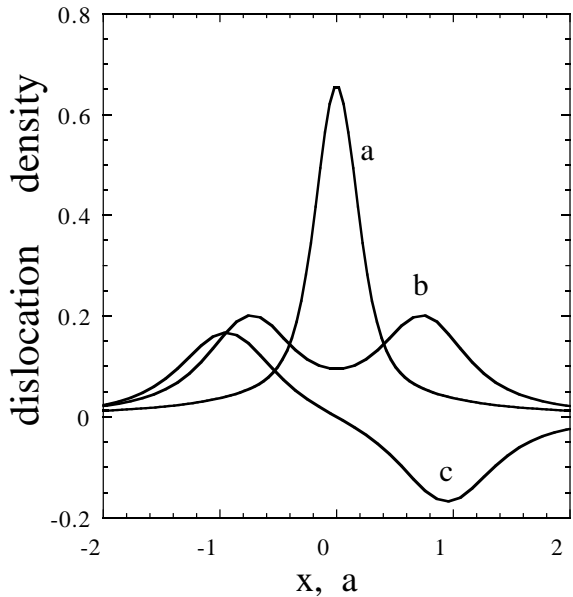,height=8.00cm}
}
\caption[a]{
{\small
Density of displacements
$ \rho(x) $ in the unit  dislocation:
a) the screw component in TiAl and for  b) the  screw and
c) edge components in CuAu as
a function of the distance (x) from the  dislocation center
(given in lattice constant units).
}
}
\label{fig:1}
\end{figure}

The results of such an analysis for a unit  dislocation
structure
in  L1$_{0}$ TiAl and CuAu
are presented in Fig. \ref{fig:1}.
As can be seen, a unit  dislocation in  TiAl is compact - in
contrast with CuAu.
 The unit dislocation structure predicted here  for TiAl (compact core)
is consistent with available experimental data \cite{tial_rev} and  
atomistic simulation results \cite{Sim2} (for an embedded atom potential that gives a 
value of the CSF energy that is similar to our ab-initio calculations) but our analysis is 
performed within a highly tractable scheme 
which allows one to  relate  qualitative differences in dislocation properties to  the
ab-initio $\gamma$-surface geometry.

\begin{figure}[!htb]
\centerline{
\psfig{figure=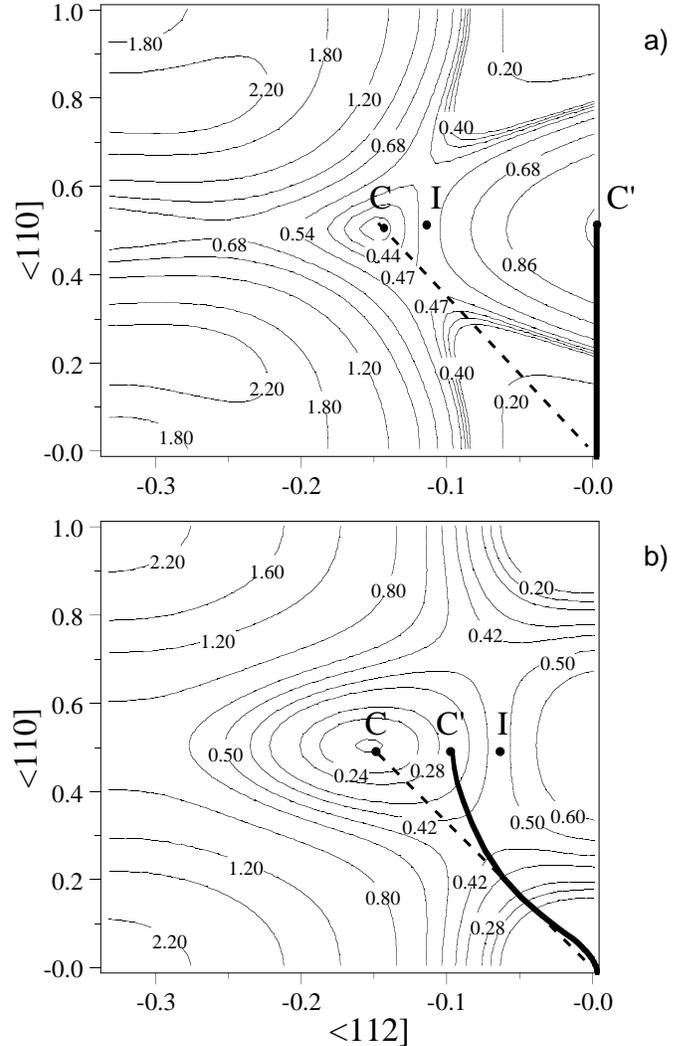,height=14.00cm}
}
\caption[a]{
{\small
Contour plots of  the $\gamma$-surface (in J/m$^{2}$ )
 for L1$_{0}$ a) TiAl and b) CuAu. The position of the  CSF is denoted by C, 
the point
 corresponding to the calculated amplitude of the edge component by C' and
 the inflection point of the corresponding section by I. The actual "dissociation path"
 (or dependence of the longitudional on lateral to the Burgers
vector components of the the displacement)
 is presented by the thick solid lines and is compared with the 
 usually assumed path (straight dashed line). 
}
}
\label{fig:2}
\end{figure}

{\it Dislocation dissociation and $\gamma$-surface geometry}.-
Focus now on the details of the $\gamma$-surface geometry which
determine  the difference  in the structure  of  TiAl and CuAu unit dislocations. 
These $\gamma$-surfaces, calculated for  the 
\{111\}  plane
with the  FLMTO method using six layer supercells and  homogeneous
periodic boundary conditions \cite{PaxtonBook,Paxton} 
without  relaxation  \cite{note2}, are presented in Fig. \ref{fig:2} as  contour plots
along with actual "dissociation" paths.
Note that the dislocation axis is taken along \mex{110} so that
displacements along \mex{112}  correspond to an edge and those
along \mex{110} have a pure screw component.
We label several special points  on  the $\gamma$-surface section
along \mex{112}  for  the  screw component u$_{1}$=0.5:
(i)  the complex stacking fault (CSF), labeled C,
(ii) the inflection  point, labeled  I, and (iii) the point  C'=(0.5,f(0.5)),
where
f(0.5) is the calculated maximal  edge component.
There is  a noticeable difference in the relative location of  I and  C
for TiAl and CuAu which results in striking differences in  the properties
of the unit  dislocations.

If the 
components of
 the restoring forces,  Eq. (\ref{ForEqn}),  along the  $u_{2}=f(u_{1})$ path  change
sign then there is more than one pole in  the
expansion,  Eq. (\ref{RhoEqn}), and so
solutions of    Eq. (\ref{eqn2dpn})  
correspond to the   
partial/fractional (split/spread ) dislocations.
In particular, this condition  is fulfilled if the corresponding $\gamma$-surface
section
has a minimum at (u$_{1}$=0.5, u$_{2}$=f(0.5)), where the
maximal edge component f(0.5) was found to be  substantially smaller than the trial
\cite{note4}
already after the first step of the iterative procedure. 
This "reduction"  is
a result of  the overlap of  the opposite sign partials displacement fields.
As a consequence of such a "reduction", C' may  be located  either
 before the  I point (CuAu) or beyond  (TiAl); this, in turn,  determines whether 
there is a minimum on the $\gamma$-surface
section and, correspondingly, whether the  dislocation is  split/spread or compact.

The iterative procedure allows us to also  find
the actual "dissociation path", curve $u_{2}=f(u_{1})$, which
is shown in Fig. \ref{fig:2} as a thick solid line.
In the case of a  spread  unit  dislocation (CuAu),  the actual path
is quite   different from the  usually assumed path\cite{note4}. However, 
 for  a compact dislocation (TiAl) the difference  of the 
 actual "dissociation path" (solid line) from  the assumed path (dashed line)  
is really  dramatic (see Fig. \ref{fig:2}).
The possibility of  such a transition from  the split/spread  to compact 
solution  (with a corresponding transformation of the "dissociation path")
has also been recently indicated by Schoeck in a 2D PN model analysis with
model type $\gamma$-surfaces   
using  the direct variational approach \cite{schoeck_98}.

The equilibrium separation distance between partials of dissociated dislocations
is traditionally determined from  the balance between elastic  (within linear elasticity theory)  
and stacking fault surface tension forces resulting in the  widely used
simple expression \cite{HirthLote,schoeck_97}.
The  PN  equations (Eq. (\ref{eqn2dpn}))  do not contain, in  an explicit form, 
the stacking fault energy
($\gamma_{f}$) which  is related  in  a very simple manner with 
the equilibrium separation distance (d) between partials of dissociated 
dislocations \cite{HirthLote}. 
However,  it is possible to demonstrate that 
this simple relation  can be recovered 
within our more general PN model approach  as the limiting case of  large separation distances.
Indeed, the energy  $\gamma_{f}$  of the fault with vector $\vec{f}$=(0.5, u$_{2}$) 
is uniquely determined  within the  local approximation through the 
two component restoring force   vector  ${\vec{ F}(\vec{u})}$  defined  for
 an arbitrary path $u_{1}=f (u_{2})$ 
as  
\begin{eqnarray}
\gamma_f = \int \limits _0^{{\vec{u}}_f}{\vec{ F}}({\vec{ u}}) d {\vec{ u}}= \int \limits
_{- \infty }^0\left(F_1\rho_1 + F_2\rho_2 \right)dx  
\label{eq:gamma_p}
\end{eqnarray}
where  $\gamma_{f}$ is uniquely defined relative to the reference point E(0)=0. 
The integral in  Eq. (\ref{eq:gamma_p}) can be evaluated analytically using solutions 
of the PN model in the form  of  Eqs. (\ref{RhoEqn}) and (\ref{ForEqn}), 
and so gives a one to one correspondence between 
$\gamma_{f}$  and parameters of the dislocation structure, 
$\gamma (\vec{u}_{f}) = \gamma ({\zeta_i}, {x_i})$.  
In the limit of large separation distances (${\zeta_i}/x_i\ll 1$),   
the following two  simplifications   
are well  justified: (i) the  relation between edge and screw components is 
close to linear ( the dependence $u_{1}=f (u_{2})$ is a straight line) and
(ii) the maximal edge component is given by the position of the CSF.
Then, 
a power series expansion of the general relation (Eq. (\ref{eq:gamma_p})) in ${\zeta _i}/x_i$ 
(up to  second order terms )  has the relatively simple form 
\begin{eqnarray}
\gamma _{f}\simeq \frac \mu {2\pi d}\left[ b_s^2 \left( 1-{{{{{{%
\left( 1-\frac 1{{\alpha}}\right) }}}}}\frac{4{\zeta}}{\pi d}}\right) -
\frac{b_e^2}{1-\nu }\left( 1-{{{{{{{\frac 8{\pi {\alpha}}}}}}}}\frac{{%
\zeta}}d}\right) \right] 
\label{eq:series}
\end{eqnarray}
where the separation distance in the limit considered  is taken as 
$d=2x_1=2x_2$ with $\alpha=\alpha_1=\alpha_2$, b$_{s}=b/2$ is the screw  component 
and b$_{e}$ the edge component of the Burgers vector and
$\alpha_{1,2}$, $\zeta_{1,2}$ are  parameters in  Eqs. (\ref{forsSEqn}) and  (\ref{forsEEqn}). 
Then it is easy to see that 
the zeroth  order terms of the $\gamma_f = \gamma ({\zeta}, d)$ 
power expansion  in ${\zeta}/d$  (for large separation distances  
 $2x_1=2x_2=d$ and ${\zeta} \ll d$)  
reproduce  exactly the well-known relation between $d$ and $\gamma_f $ \cite{HirthLote}.
%
This simple estimate, however, 
becomes  obviously    inaccurate for  dislocations with small
 separation distances, as in the  case of  CuAu and TiAl considered here.
This  result is non-trivial since it  allows one  to relate  rigorously the  PN model 
and  often used  simple  elasticity \cite{HirthLote}   descriptions. 
We also find   that  the higher order (in $\zeta/d$) contributions in Eq. \ref{eq:series} have  negative sign,
thus the simple relation  $\gamma \sim  1/d $  corresponds to an upper  estimate 
which becomes  increasingly inaccurate for  dislocations with small
 separation distances, d.
We should emphasize%
, however, 
that
in accordance with our 2D PN analysis 
the CSF energy alone is not a sufficient characteristic  to describe
important details of the dislocation structure.
Hence,  additional parameters of the entire
$\gamma$-surface (see Eq. \ref{eq:series})  have to be taken into account
for a correct analysis - which  is especially  important in the case of small
dislocation spreading  
(as found in our analysis for $\gamma$-surfaces calculated for TiAl and  CuAu). 

{\it The Peierls stress}. - Consider now how such a characteristic of the
dislocation mobility as
the Peierls stress,   $\sigma _P$,  can be determined within the  proposed
PN model analysis scheme.
According to the  approximations introduced by Nabarro \cite{Nabarro}
(which often work reasonably well \cite{Nabarro97}), 
$\sigma _P$  is defined as 
\begin{eqnarray}
& \sigma_{p}=\frac 1b\left[ \frac{d\Phi{_P(l)}}{dl}\right]_{\max }&,
\label{sigma}
\end{eqnarray}
through the  so-called  ``misfit energy'' per unit length of dislocation, $\Phi
_P(l) $,
which can be written in the form \cite{ourPN1}
\begin{eqnarray}
\Phi _P(l)  = \Phi _P^{(0)} & + & \frac h 2 \sum_{s=1} J(s)
\left[\cos (2\pi sl/h) + \right.
\nonumber \\
& + &
\left.
\cos (2\pi s(l-\delta )/h)\right]
\label{Phi2}
\end{eqnarray}
where
\begin{eqnarray}
J(s)=\frac{-1}{2\pi is}\int\limits_{-\infty }^\infty
\vec{F} \vec{\rho} \exp (2\pi isx/h)dx
\label{JEqn}
\end{eqnarray}
Here $h$ is the  lattice period in a direction perpendicular to
the  dislocation line, \cite{Duesbery89},
and  $\delta$ is the relative displacement of the atomic
arrays in neighbouring
\{111\} planes of the fcc lattice.
Hence,  the   determination  of $\sigma _p$ is reduced to the integration of
the product of analytic vector functions ( Eqs. (\ref{RhoEqn}) and
(~\ref{ForEqn})).

We used these expressions to calculate Peierls stresses for the unit
dislocations in TiAl and CuAu and found for the screw orientation
that  the ratio $\sigma_{p}/\mu$ (here $\mu$ is the shear modulus)
for CuAu (0.0005)  is significantly smaller than for TiAl (0.024).
This difference is attributed to the  spreading  of the unit  dislocation in
CuAu, since    $\sigma_{p}/\mu$ for the  unspread core solution
 is about the same as for TiAl.
 This result is quite remarkable  since  it clearly demonstrates
the importance of the entire $\gamma$-surface geometrical parameters as 
an integrated characteristic of the
interatomic interactions of solids (including chemical bonding contributions)
    in the context of the analysis of dislocation properties.

We  emphasize here the importance 
of such  characteristics that are naturally revealed in our  2D PN model analysis  as  a 
"dissociation path" for 
the analysis  of the fracture behaviour. 
As is well-known in the scope of Rice-Thomson (R-T) criteria, crack propagation is described 
by  the competition of the crack opening (characterized by  the surface energy $\gamma_{s}$)
and dislocation emission (characterized by unstable stacking fault energy $\gamma_{us}$)
processes. An analysis of this competing process  arrives at the criterion 
of brittle crack propagation,  $\gamma_{s}/\gamma_{us} > $ 2.9\cite{JRRice:92,thomson-book},   
which  despite its simplicity captures 
the general tendencies  in  materials fracture behaviour \cite{thomson-book}. 
Dislocation emission from the crack tip within the R-T criteria is
characterized by the energy barrier, $\gamma_{us}$.    
This energy barrier  is usually estimated assuming a  straight line "dissociation"  path\cite{note4}.
As we demonstrated above, the actual "dissociation"  path may  deviate significantly 
from an assumed straight line. In cases of spread dislocations (CuAu),  corresponding 
changes in    $\gamma_{us}$  estimates are  likely to be small  but  may be quite significant  in  
cases of compact  dislocation cores (TiAl). 
In fact, if  features of the actual "dissocation" path are taken into account
to determine   $\gamma_{us}$ for  TiAl,  the R-T criterion ($\gamma_{s}/\gamma_{us}=1.7$) predicts 
brittle  crack propagation - opposite  to  the  result which 
one  obtains assuming a straight "dissocation" path ($\gamma_{s}/\gamma_{us}=4.5$).  

\section{Conclusions}

 We proposed a general and physically transparent scheme
for analyzing  the 2D PN model  with a general restoring force law.
We provided realistic  examples (TiAl, CuAu)
and demonstrated that  
%
dislocation properties can be predicted and explained 
on the basis  of ab-initio calculations of the $\gamma$-surfaces.
In particular, the actual "dissociation path" was  determined  and proved to be 
an essential feature of the 2D PN model  and an important characteristic for
understanding   dislocation dissociation processes  and related mechanical
properties (tendencies in fracture behaviour).
We showed that differences in  
dislocation properties (compact vs. spread) for realistic (complex) situations as  in TiAl 
and CuAu can be explained from  the ab-initio $\gamma$-surface geometry.
%
This demonstrates the advantages  of the proposed 
approach which establishes, within the 2D PN model,  
explicit   relations 
 between $\gamma$-surface geometry parameters
(which can be accurately  determined using  ab-initio techniques) 
and dislocation structure and mobility
 (processes  that approach mesoscopic scales).

\section{Acknowledgements}

We are grateful to D. Dimiduk, C. Woodward and S. Rao for useful
and encouraging discussions, and for providing preprints prior to publication.
We also grateful to D. Novikov and M. Schilfgaarde for
help with the FLMTO code.
This work was supported by the Air Force Office of Scientific Research
(Grant No. F49620-95-1-0189) and  grant
of computer time
 at the Pittsburgh Supercomputing Center (supported by the NSF).


\begin{thebibliography}{10}
\bibitem{note1}
Note that this simple  (decoupled) structure  of the 
2D PN model equations is valid not only for isotropic but also for anisotropic elastic media
if the  dislocation axis lies along a twofold symmetry axis
\cite{seeger_53} in which case  we use  an 
anisotropic modulas $\mu_{\vec{n}\vec{b}}$ corresponding to the shear
in  the slip plane $\vec{n}$  along  $\vec{b}$.
\bibitem{note3}
The limiting case corresponding to the  simple sinusoidal restoring force law
can be easily obtained from solutions of Eqs. (\ref{disSEqn} - \ref{forsEEqn})
(first two terms for $\alpha$=1).
Note here, that only in this simple case ($\alpha$=1)  can solutions 
of the 2D PN model  $ u_{\beta} (x) $ be  constructed as  superpositions
of the simple  function  being solutions of the 1D PN model with sinusoidal restoring force
law. This  type of simple trial fuction  was used  by Schoeck in a variational 
scheme for solving the 2D PN model \cite{schoeck_94,schoeck_98}
\bibitem{note4}
In contrast with the screw, the  maximal value
of the edge component is unknown.
Usually, it is assumed to be determined by the
position of the  CSF which may be justified only in the limit of
large splitting distances.
In this limit,  "dissociation path" may be well approximated 
by a straight line which is used only as a trial in our iterative
procedure for  determining the  actual dependence of  the screw component of displacements on  the edge.
\bibitem{note2}
Strictly speaking, the effects of the relaxation  on the  $\gamma$-surface
energetics goes beyond the "original" PN model
concept and  so are not considered in this
work, which is aimed at  studying the  model within the  original formulation.
\end{thebibliography}



\end{document}